\documentclass[sigconf,screen]{acmart}

\setcopyright{none}
\renewcommand\footnotetextcopyrightpermission[1]{}
\AtBeginDocument{%
  }

\acmConference[WWW '26]{Proceedings of the ACM Web Conference 2026}{April 13--17, 2026}{Dubai, UAE}
\acmBooktitle{Proceedings of the ACM Web Conference 2026 (WWW '26), April 13--17, 2026, Dubai, UAE}

\setcopyright{acmlicensed}
\copyrightyear{2026}
\acmYear{2026}
\acmDOI{XXXXXXX.XXXXXXX}
\acmISBN{978-1-4503-XXXX-X/24/06}

\usepackage{pifont}
\usepackage{tcolorbox}
\usepackage{diagbox}
\usepackage{enumitem}
\usepackage{booktabs}
\usepackage{xcolor}
\usepackage{hyperref}
\usepackage{graphicx}  % 用于调整圆圈大小
\usepackage{array}
\usepackage{caption}
\usepackage{subcaption}
\usepackage{multirow}
\usepackage{tikz}
\usepackage{siunitx}
\usetikzlibrary{arrows.meta, shapes, positioning}
\definecolor{mygreen}{RGB}{0,200,170}
\definecolor{mypink}{RGB}{250,50,150}
\definecolor{myblue}{RGB}{50,150,250}
\definecolor{myyellow}{RGB}{250,240,50}
\definecolor{deepgreen}{RGB}{31,130,31}   
\definecolor{deepblue}{RGB}{16,73,124}  

\definecolor{myorange}{HTML}{b35809}
\setcopyright{none}
\renewcommand\footnotetextcopyrightpermission[1]{}

\begin{document}
% \title{Polaris: MIR-Centric Binary Obfuscation as an LLVM Backend Framework}

% \title{Understanding JavaScript Engine Vulnerabilities and Exploitation Mechanisms: An Empirical Study}

\title{When Flaws Cascade: Understanding Vulnerabilities and Exploitation Chains in JavaScript Engines}

% \title{Understanding Vulnerabilities and Exploitation Mechanisms in Web Browser Engines}

\author{Yuhan Ma}
\orcid{0009-0008-6625-7446}
\affiliation{%
  \institution{Tianjin University}
  \city{Tianjin}
  \country{China}
}
\email{mayuhan@tju.edu.cn}

\author{Jiongchi Yu}
% \correspondingauthor
\orcid{0000-0002-2888-4499}
\affiliation{%
  \institution{Nanyang Technological University}
  \city{Singapore}
  \country{Singapore}
}
\email{jiongchiyu@acm.org}

\author{Xiaofei Xie}
\orcid{0000-0002-1288-6502}
\affiliation{%
  \institution{Singapore Management University}
  \city{Singapore}
  \country{Singapore}
}
\email{xfxie@smu.edu.sg}

\author{Qiang Hu}
\orcid{0000-0002-8251-1669}
\affiliation{%
  \institution{Tianjin University}
  \city{Tianjin}
  \country{China}
}
\email{qianghu@tju.edu.cn}

\author{Zhiyi Zhang}
% \orcid{0000-0002-8251-1669}
\affiliation{%
  \institution{Qi An Xin Group Corp}
  \city{Beijing}
  \country{China}
}
\email{hi.zhiyi@qq.com}

\author{Junjie Wang}
\orcid{0009-0002-3847-6760}
\affiliation{%
  \institution{Tianjin University}
  \city{Tianjin}
  \country{China}
}
\email{junjie.wang@tju.edu.cn}

%\author{\IEEEauthorblockN{Ang Zhou\textsuperscript{1}, Yuhan Hu\textsuperscript{1}, Gao Yang\textsuperscript{1}, Junsong Fu\textsuperscript{1,*}\thanks{* Junsong Fu is the corresponding author.} and Hongye Cai\textsuperscript{2}}
%\IEEEauthorblockA{
%\textsuperscript{1}Beijing University of  Posts and Telecommunications \\\textsuperscript{2}First Research Institute of the Ministry of Public Security of PRC \\
%\{r1mao233, huyuhan, yang\_gao, fujs\}@bupt.edu.cn, caihongye@163.com}
%}

\begin{abstract}
    JavaScript engines are pivotal to modern web browsers, enabling the execution of dynamic and interactive web applications. However, their complexity and widespread adoption make them prime targets for attackers exploiting vulnerabilities. While existing research has focused on detecting vulnerabilities of JavaScript engines, a significant gap remains in systematically understanding the characteristics of these vulnerabilities, including their symptoms, root causes, and exploitability. This paper bridges this gap by presenting the first comprehensive empirical study on vulnerabilities in JavaScript engines, investigating their characteristics and potential exploitation strategies.

We construct a dataset comprising 241 vulnerabilities across four mainstream JavaScript engines from 2017 to 2024. Through in-depth analysis, we first develop taxonomies for symptoms and root causes. Building on this understanding, we investigate the exploitability of these vulnerabilities, identifying key prerequisites and extracting vulnerability trigger chains that demonstrate how logical errors propagate into memory safety violations. Additionally, we analyze the mitigation strategies to counter these exploitations. Finally, we summarize key implications for various stakeholders, including developers and researchers, offering actionable insights to improve the security and resilience of JavaScript engines.
\end{abstract}



\maketitle

\section{Introduction}

JavaScript~(JS) is one of the most widely used programming languages, forming the backbone of interactive and dynamic web applications. According to recent statistics~\cite{W3Techs2024JavaScriptUsage}, JS is utilized on the client side by 99\% of all websites. Virtually every modern web browser integrates a JS engine, such as Google’s V8 in Chrome, JavaScriptCore~(JSC) in Safari, SpiderMonkey~(SM) in Firefox, and ChakraCore~(CH) in Microsoft Edge (formerly used in Internet Explorer), underscoring its critical role in delivering rich and responsive user experiences. Beyond web browsers, JS has evolved into a versatile language, extending its applications to mobile app development and desktop applications, and its engine serves as the key component for executing code written in JS language.

\textbf{Motivation.} JS remains the dominant language for client-side web development. Its dynamic nature and runtime flexibility enable rich web experiences but also introduce complexity in engine implementation. Features such as dynamic typing, prototype-based inheritance, and runtime code execution (e.g., via \texttt{eval}) complicate the development of secure and efficient JS engines. While similar constructs exist in other languages (e.g., \texttt{exec} in Python), the ubiquity of JS in browsers and its exposure to untrusted input elevate the risk profile of JS engines in real-world scenarios.

Modern JS engines are complex systems that integrate multiple subsystems, including interpreters, just-in-time (JIT) compilers, and garbage collectors. Their performance-critical nature and tight coupling with web browsers make them attractive targets for attackers. Existing studies~\cite{Lim2021, Hawkes2019} have highlighted the prevalence of vulnerabilities in JS engines, with many high-impact CVEs demonstrating their security relevance.
Real-world incidents further underscore this concern. For instance, the 2016 NSO Group spyware attack on UAE activist Ahmed Mansoor exploited a Safari JS engine vulnerability (CVE-2016-4657~\cite{CVE-2016-4657}) as part of a zero-click jailbreak chain~\cite{CitizenLab2016}. While such vulnerabilities are not unique to JS engines, their position as the gateway to executing client-side logic on the web amplifies their impact.

Despite increasing research on fuzzing and testing techniques for JS engines~\cite{holler2012langfuzz, wang2019superion, gross2018fuzzil, wang2023fuzzjit}, and some efforts to examine specific classes of vulnerabilities (e.g., JIT bugs~\cite{li2019jit, gross2018jit}), the landscape remains fragmented. Prior work often focuses on isolated cases or lacks a systematic, security-centered taxonomy. Moreover, publicly available vulnerability databases~\cite{js-vuln-db, kang2021review} provide limited insights into exploitability or root causes. This gap motivates our effort to empirically study JS engine vulnerabilities through the lens of both security impact and root cause taxonomy.

To address this gap, we conduct the first empirical study aimed at understanding vulnerabilities in JS engines. This study seeks to answer three key research questions: \textbf{RQ1}: What are the symptoms of vulnerabilities in JS engines? \textbf{RQ2}: What are the prevalent root causes and distribution characteristics of these vulnerabilities? \textbf{RQ3}: How can these vulnerabilities be used for exploitation?

\textbf{Contributions.}
To address these research questions, we first build a comprehensive dataset of JS engine vulnerabilities collected from multiple sources, focusing on the four major engines: V8, JSC, SM, and CH. After filtering, 241 vulnerabilities are retained for analyzing their characteristics, and 22 representative trigger chains are manually constructed and analyzed.
% \textit{understanding the characteristics of these vulnerabilities and examining how they can be exploited and mitigated.}
% Our empirical study, based on this curated dataset, 
%Analyzing the distribution of vulnerabilities by location highlights the components or phases most susceptible to security issues, guiding researchers and developers to prioritize critical areas for improvement.
Regarding vulnerability characteristics, our study mainly examines the symptoms and their root causes. Identifying symptoms helps understand their impact and facilitates the design of targeted detection strategies tailored to specific manifestations. Investigating root causes provides valuable insights into the underlying issues, enabling the development of robust preventative measures to enhance the overall security of JS engines.
% For exploits, our focus is on understanding these vulnerabilities from both attack and defense perspectives—how they can be leveraged in attacks and how they can be mitigated to counter such threats. 
% Specifically, we identify vulnerability trigger chains that represent the flows from the root cause to the memory corruption errors. 
% By analyzing the characteristics of these chains, we can gain insights into the impact of 
% vulnerabilities and understand strategies for their mitigation.
For exploits, we analyze these vulnerabilities from both attack and defense perspectives, aiming to understand how they are leveraged in attacks and how they can be mitigated. Specifically, we identify vulnerability trigger chains that trace the flow from root causes to memory corruption. Analyzing these chains reveals the impact of vulnerabilities and informs effective mitigation strategies.

% concentrate on the process that leads to memory corruption errors, as they are relatively easier to exploit. By studying attacks and mitigation techniques, we aim to gain insights into the impact of vulnerabilities and understand strategies for their mitigation and repair.

In summary, we identify 6 types of symptoms and 15 types of root causes. 
Among the symptoms, the majority fall into two main categories: crash (70.54\%) and erroneous functionality (27.39\%). 
For root causes, most vulnerabilities originate from incorrect type handling (31.54\%) and incorrect data structure handling (29.46\%), both of which are closely tied to the dynamic and weakly typed nature of JS. 
Furthermore, issues related to engine optimization algorithms contribute significantly, representing 18.67\% of the root causes.
Moreover, we summarize various vulnerability trigger chains to illustrate how logical errors in the engine propagate across multiple stages, ultimately leading to memory safety violations. This analysis provides insights into exploitation and identifies critical points for effective mitigation. Based on our findings, we further offer key implications through in-depth discussions for various stakeholders.

In summary, this paper makes the following contributions:
\begin{itemize}[leftmargin=*] 
\item We construct a comprehensive dataset containing 241 vulnerabilities across four mainstream JS engines, covering a seven-year timespan since Jan 1st, 2017. 
% The detailed data can be found on our anonymous website~\cite{website}.

% \item We conduct the first empirical study to systematically investigate the characteristics of vulnerabilities in JS engines. This includes detailed statistics on the distribution of vulnerabilities and the taxonomies of vulnerability symptoms and root causes. %, providing valuable insights into the nature of JS engine vulnerabilities.
\item We conduct the first empirical study of JS engine vulnerabilities, providing statistics on their distribution and taxonomies of symptoms and root causes.

% \item We analyze the potential exploitability of vulnerabilities in JS engines and construct vulnerability trigger chains to help better understand security risks in JS engines. Additionally, we explore mitigation measures in countering these exploitations.
\item We analyze JS engine vulnerability exploitability, construct trigger chains to understand risks, and mitigation strategies.

\item We provide key implications for various stakeholders, offering actionable insights to enhance the security of JS engines.

\end{itemize}
\section{Background}
\label{sec:background}

% \textcolor{red}{{subtitle?}}

The primary role of a JS engine is to transform JS source code into executable code and run it efficiently at runtime. 
Unlike traditional compilers, which compile static languages (e.g., C/C++ or Java) into machine code in a single pass, JS engines must dynamically handle JS's flexible and interpreted nature. Instead of precompiling the entire program, they process plain-text JS at runtime, translating it into bytecode, which is then executed either by an interpreter or a JIT compiler. 
To balance fast startup and runtime performance, JS engines use a hybrid execution model, 
% initially 
interpreting code for immediate execution and progressively compiling frequently executed portions into optimized machine code through JIT compilation.

\begin{figure}[t]
\centering
\includegraphics[width=0.8\linewidth]{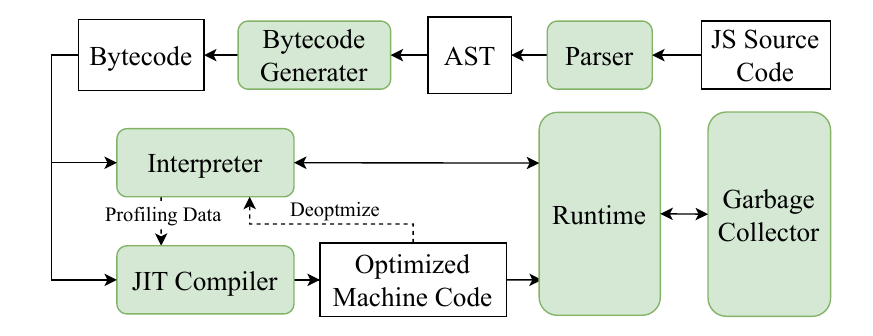}
\captionsetup{justification=centering}
\caption{The architecture of the JS engine.}
% \vspace{-15pt}
\label{fig:pipeline}
\end{figure}

\autoref{fig:pipeline} shows the general architecture of JS engines, which consists of six main components: Parser, Bytecode Generator, Interpreter, Runtime, Garbage Collector, and JIT Compiler. These components work together to support the JS execution process, which is structured into multiple phases, each handling specific tasks. For example:

\begin{itemize}[leftmargin=*]
    \item \textbf{Parsing \& Bytecode Generation}: The engine constructs the AST and converts it into bytecode to support JS execution.
    \item \textbf{Runtime Execution}: Code execution begins with interpretation, allowing immediate execution, while frequently used code is progressively optimized through JIT compilation.
    \item \textbf{JIT Optimization}: The JIT compiler improves execution by speculatively making assumptions about runtime behavior and applying optimizations. 
    Common speculation techniques include Type Speculation phase  and Range Speculation phase. 
    Common optimizations include phases such as Constant Folding (CF), Common Subexpression Elimination (CSE), Reduction, Escape Analysis, Lowering, Dead Code Elimination (DCE), Bounds Check Elimination (BCE), Instruction Scheduling, Type Check Elimination (TCE), and Write Barrier Elision (WBE).

    % Common speculation techniques include \textbf{Type Speculation}, which predicts the likely types of variables, and \textbf{Range Speculation}, which assumes constraints on numerical values to enable more aggressive optimizations.
    
    \item \textbf{Garbage Collection}: The Garbage Collector periodically reclaims memory via marking and sweeping.
\end{itemize}

\begin{figure*}[t]
\centering
\includegraphics[width=0.8\linewidth]{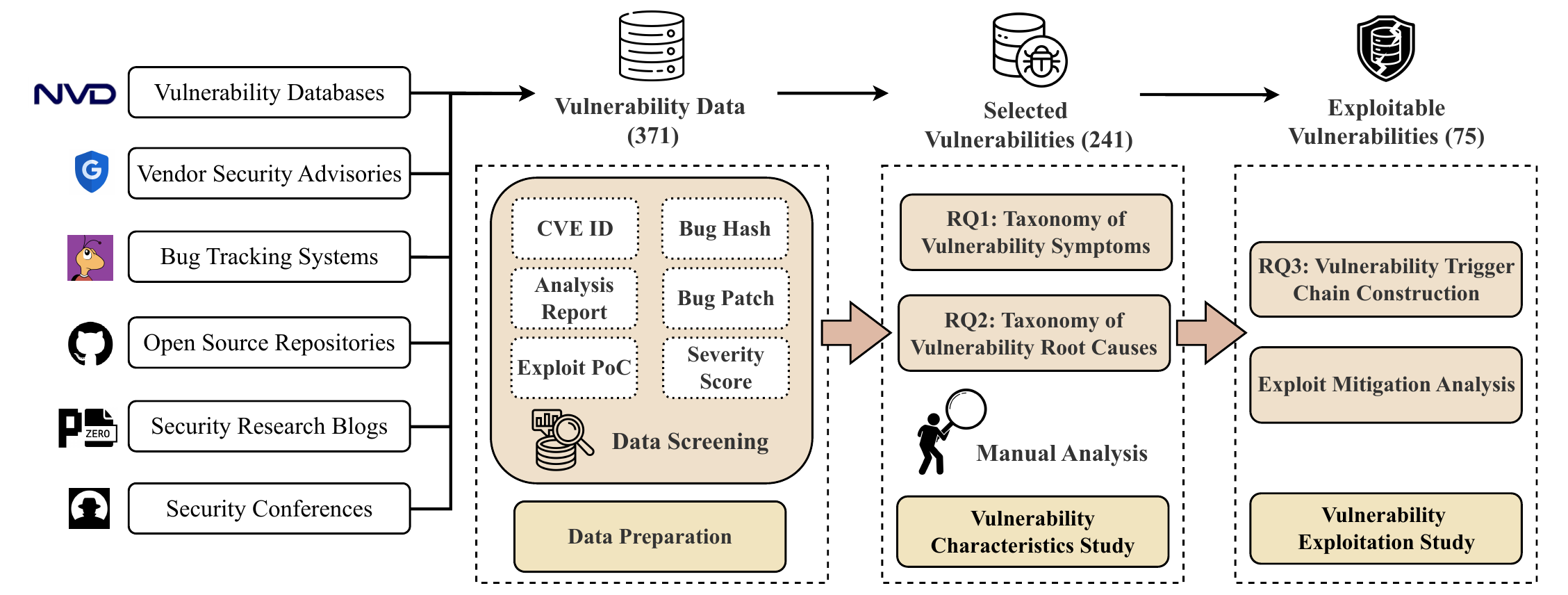}
\captionsetup{justification=centering}
\caption{The overview of our study.}
\label{fig:overview}
\end{figure*}

\section{Methodology}

\autoref{fig:overview} outlines the workflow of our study. We collect all publicly reported vulnerabilities from the four major JS engines, filter for high-impact, well-documented cases, and build a dual taxonomy of symptoms and root causes. We then trace each vulnerability across engine phases to reconstruct its exploitation chain and assess the mitigations that disrupt it.

\subsection{Data Preparation}

We focus on the four most widely deployed JS engines: \textit{V8}, \textit{JSC}, \textit{SM}, and \textit{CH}, which underpin major browsers and runtimes. To build a comprehensive dataset, we aggregate all publicly reported engine vulnerabilities from diverse sources, including CVE databases, official security advisories, bug-tracking systems, research blogs, and conference disclosures. In total, we identify \textbf{1,215} candidate entries before screening.

We then filter out outdated or insufficiently documented records, retaining only well-analyzed vulnerabilities published since 2017. To ensure analytical depth, we prioritize high-severity cases (CVSS > 7.0) or those with bounty recognition, yielding \textbf{241} high-impact vulnerabilities. For each case, we collect patch and discussion data from the corresponding bug trackers to reconstruct resolution context and technical reasoning. Among these, \textbf{75} vulnerabilities include exploitable proofs of concept (PoCs), which we further analyze to trace exploitation trigger chains. Detailed data collection and triage procedures, including keyword selection and supplementary mining from restricted trackers, are provided in Appendix~\ref{appendix:datacollection}.

\subsection{Manual Analysis}

% Drawing on the gathered \textbf{241} well-documented vulnerabilities, we first conduct a taxonomy of their symptoms and root causes to better understand the characteristics of these vulnerabilities. Subsequently, we reconstruct their exploitation trigger chains to study the attacking behaviors.
% and explore the patching strategies to defend against successful exploitation.
To systematically understand JS engine vulnerabilities, we manually analyze the 241 filtered samples across two dimensions: (1) their surface-level symptoms and root causes, and (2) their exploitability and propagation behavior.

% \noindent
\textbf{Vulnerability Characteristics.} 
We adopt open coding~\cite{shen2021comprehensive,quan2022towards} to extract taxonomies for symptoms and root causes. The dataset is split into two halves and annotated in two phases. Three security researchers independently annotate the first half. Each has 3–5 years of experience in software security, with prior work involving vulnerability analysis and fuzzing for JS engines. 
% They have co-authored peer-reviewed publications in top-tier security venues and are part of the research team conducting this study. 
After the initial annotation, the three analysts reconciled their labels. Unresolved cases were adjudicated by a senior team member with over 10 years of vulnerability research experience and multiple top-tier publications, who served as the final arbiter when consensus could not be reached.
% In the study, we follow previous studies~\cite{shen2021comprehensive,quan2022towards,yu2024bugs} by utilizing the open coding technique to identify the symptoms and root causes associated with vulnerabilities. We split the gathered \textbf{241} vulnerabilities into two halves, undertaking a dual-phase annotation strategy for our analysis. 
% Initially, three analysts with professional backgrounds in software security independently verify and categorize half of the vulnerability samples. Each analyst has 3–5 years of research or engineering experience in areas including vulnerability analysis, program analysis, and JavaScript engine internals. All three have participated in academic or industrial security projects, with publications in venues, and prior experience in analyzing security patches and root causes of real-world bugs. These analysts classify the vulnerabilities according to distinct categories of symptoms and root causes.
% After the initial annotation phase, all three analysts engage in a group discussion to review the preliminary taxonomy results. Any disagreements are resolved by a senior security expert with over ten years of experience in vulnerability research, who has contributed to multiple studies on JS engine vulnerabilities and has published in top-tier venues. This expert serves as the arbitrator, leading the discussion and determining the final categorization decisions when consensus cannot be reached.
The analysis proceeds similarly with the second half of the data. Following discussions on the preliminary taxonomy, the three experts reassess and label the remaining half of the vulnerabilities. To evaluate the consistency of the categorizations made, we employ Cohen’s Kappa (k) coefficient~\cite{hsu2003interrater} to measure inter-rater reliability, where higher values indicate greater agreement. After completing the annotation of the entire dataset, we observed that in the first round of analysis, the inter-rater reliability (k) reaches \textbf{0.51}, indicating moderate agreement. After the discussion and the second labeling phase, the k value rose to \textbf{0.85}, demonstrating almost perfect agreement.

% \noindent
% \textbf{Vulnerability Exploitation.} For the vulnerability trigger chain construction, four security experts manually analyze the 75 exploitable vulnerabilities based on the execution of the PoCs. 
% % The detailed methodology is introduced in Section~\ref{}. 
% Specifically, we trace the exploitation pathways starting from the root cause of each vulnerability, following the function calling chain through each execution phase~\footnote{Due to the limit of space, we provide a detailed illustration of different phases for all JS engines on our website~\cite{website}.} within the JavaScript engines until the reported final symptom. 

% In total, we extract 22 triggering chains, as detailed in Section~\ref{sec:vulnerabilitytrigger}.

\textbf{Vulnerability Exploitation.} 
We concentrate on the 75 vulnerabilities that ship with a reproducible PoC or a publicly documented exploit chain. This subset offers the concrete runtime evidence needed to follow a bug’s propagation across engine phases. Four members of our team replayed each PoC, tracing the call stack from the root cause through every execution phase to the final symptom. The process revealed 22 distinct trigger-chain patterns, which we detail in \autoref{sec:vulnerabilitytrigger}.

\section{Symptom Taxonomy}\label{sec:symptom}

Through manual analysis, we have identified two main categories of vulnerability symptoms in JS engines. We observe that these symptoms bear similarities to those discussed in traditional software bug studies~\cite{quan2022towards,shen2021comprehensive}. Therefore, here we focus primarily on those symptoms specific to JS engines, as detailed in the symptom distribution shown in \autoref{tab:sd}, and we classify five vulnerabilities as \textit{Others} due to ambiguous symptom information that could not be clearly categorized. 

% We put a more comprehensive representation and discussion of these symptoms on our website~\cite{website}.

% across various JS engines, categorized by their symptoms. 

% The remaining symptom categories are detailed as follows:

% \fei{can mention that many symptoms are similar to traditional ones, we ignore their details and mainly introduce the JS engine-specific ones? A more detailed symptom representation and introduction can be found on the website.}

% All the symptoms of JS engines are presented as follows: 
\begin{table}[t]
\renewcommand\arraystretch{0.5}
    \centering
    \footnotesize
    \caption{Distribution of Symptoms}
    \resizebox{0.4\textwidth}{!}{
    \begin{tabular}{c|c|c|cccccc}
        \toprule
        \multicolumn{3}{c|}{\textbf{Symptom}} & \textbf{V8} & \textbf{JSC} & \textbf{SM} & \textbf{CH} & \textbf{Total} \\
        \midrule
        \multirow{5}{*}{\textbf{Crash}} 
        & \multirow{3}{*}{\textbf{MSV}} & \textbf{OOB}  & 25 & 16 & 2 & 6 & 49 \\
        \cmidrule{3-8}
        &  & \textbf{TC}  & 15 & 7 & 4 & 41 & 67 \\
        \cmidrule{3-8}
        &  & \textbf{UAF} & 4 & 5 & 11 & 5 & 25 \\
        \cmidrule{2-8}
        & \multicolumn{2}{c|}{\textbf{VF}} & 19 & 2 & 7 & 1 & 29 \\
% \cmidrule{2-8}
        % & \multicolumn{2}{c|}{\textbf{Total\textsubscript{crash}}} & 63 & 30 & 24 & 53 & 170 \\
        \midrule
        \multicolumn{3}{c|}{\textbf{EF}} & 30 & 15 & 5 & 16 & 66 \\
        \midrule
        \multicolumn{3}{c|}{\textbf{Others}} & 0 & 1 & 0 & 4 & 5 \\
        \midrule
        \multicolumn{3}{c|}{\textbf{Total}} & 93 & 46 & 29 & 73 & 241 \\
        \bottomrule
    \end{tabular}
    }
    \label{tab:sd}
\end{table}
% \vspace{-5pt}

% \paragraph{\textbf{Crash}} 
\noindent\textbf{\hypertarget{Crash}{Crash}}
occurs when a JS engine terminates unexpectedly, accounting for most of the vulnerabilities (70.54\%) we studied. These can be further categorized into \textbf{Memory Safety Violation (MSV)} and \textbf{Validation Failure (VF)}. 
% MSV occurs when the engine improperly accesses memory (e.g., writing beyond allocated boundaries), which ultimately forces the engine to exit unexpectedly. These violations encompass three categories, including Out-of-Bounds (OOB) access, Type Confusion (TC), and Use-After-Free (UAF), each of which can pose severe security risks. 
MSV occurs when the engine improperly accesses memory (e.g., writes beyond allocated bounds), causing unexpected termination. These violations include Out-of-Bounds (OOB) access, Type Confusion (TC), and Use-After-Free (UAF), each posing serious security risks.
VF occurs when the program’s state diverges from predefined assertions and debugging utilities, which are employed to identify breaches of expected invariants or assumptions in the code.
% VF occurs when the program’s state diverges from the predefined assertions and custom debugging utilities by developers, where these assertions are employed to identify breaches of expected invariants or assumptions embedded within the code.

% Crash is the most common symptom across all four JS engines, with V8 (63), JavaScriptCore (30), SpiderMonkey (24), and ChakraCore (53) being affected. 
% Among these crashes, MSV accounts for the majority (82.94\%) across all four JS engines. We observe that different JS engines are prone to specific types of vulnerabilities. Specifically, V8 predominantly exhibits OOB symptoms at 56.82\%, along with a substantial incidence of TC at 34.09\%. JSC displays similar patterns to V8, with OOB and TC distributions at 57.14\% and 25.00\%, respectively.  SM, however, experiences fewer overall crashes but has the highest incidence of UAF at 64.71\%, likely due to complexities in its GC subsystem. For CH, with TC symptoms accounting for 78.85\%, this highlights its deficiencies in type management.
Among these crashes, MSVs account for the majority (82.94\%) across all four JS engines. Different engines show distinct vulnerability tendencies: V8 primarily exhibits OOB (56.82\%) and TC (34.09\%) symptoms, while JSC displays similar distributions (57.14\% and 25.00\%). SM shows fewer crashes overall but has the highest UAF rate (64.71\%), likely due to its complex GC subsystem. In contrast, CH has a dominant TC symptom rate of 78.85\%, reflecting weaknesses in its type management.
VF is more prevalent in V8 and SM, accounting for 20.43\% and 24.14\% of their vulnerabilities, respectively. This prevalence is mainly because V8 and SM implement stricter assertion checks in debug mode to detect logical inconsistencies early on. Notably, 65.5\% of VF can escalate into MSVs in release versions, emphasizing the importance of secured validation, especially for memory-related code.
% as part of vulnerability detection and mitigation strategies across JS engines.

\noindent
\begin{tcolorbox}[colback=gray!15, colframe=black,boxrule=1pt,before=\vspace{5pt},after=\vspace{0.5pt},left=1pt,right=1pt,top=0.5pt,bottom=0.5pt]
\textit{\textbf{Finding 1}: Crashes are the most common symptom (70.54\%), predominantly manifested as MSVs (82.94\%) across all studied JS engine vulnerabilities. The high prevalence of VF in V8 and SM indicates their stricter assertion checks, which serve as early indicators of potential security issues.}
\end{tcolorbox}

% \paragraph{\textbf{Erroneous Functionality (EF)}} \label{par:ef}
\noindent\textbf{\hypertarget{EF}{Erroneous Functionality (EF)}} appears when a JS engine behaves unexpectedly without crashing. For example, aggressive optimization techniques may introduce errors by making invalid assumptions about variable types, which leads to inaccurate results, like CVE-2018-0777~\cite{CVE-2018-0777}. We observe that EF is a relatively common symptom, accounting for 27.39\% of the studied vulnerabilities.
% 32.3\% of the vulnerabilities in V8, 32.6\% in JSC, 17.2\% in SM, and 21.9\% in CH.
% Unlike crashes, EF vulnerabilities are more subtle and often manifest as incorrect computations or optimization errors. Detecting these vulnerabilities is significantly more challenging than identifying crashes because they generally do not trigger runtime errors. Instead, EF vulnerabilities remain latent during execution and only become apparent under specific inputs or execution paths.
Unlike crashes, EF vulnerabilities are more subtle, often appearing as incorrect computations or optimization errors. They are harder to detect since they rarely trigger runtime errors, remaining latent during execution and surfacing only under specific inputs or paths.
% To improve the detection and mitigation of such vulnerabilities, more advanced testing methods are required, including semantic analysis, differential testing, and large-model-based error pattern recognition.

\noindent
\begin{tcolorbox}[colback=gray!15, colframe=black,boxrule=1pt,before=\vspace{1pt},after={\vspace{0pt}},left=1pt,right=1pt,top=0.5pt,bottom=0.5pt]
\textit{\textbf{Finding 2}: EF vulnerabilities are more difficult to observe compared to crashes, yet they account for an average of 27.39\% vulnerabilities.}
\end{tcolorbox}

\section{Root Cause Analysis}
% \section{Root Cause Taxonomy}
\label{sec:rootcause}
\subsection{Root Cause Taxonomy}
We further analyze the root causes of vulnerabilities in JS engines, as illustrated in \autoref{fig:rc}. We identify 6 main categories and divide them into \textcolor{black}{15} leaf categories based on their execution mechanisms:

\begin{figure*}[t]
\centering
\includegraphics[width=0.75\linewidth]{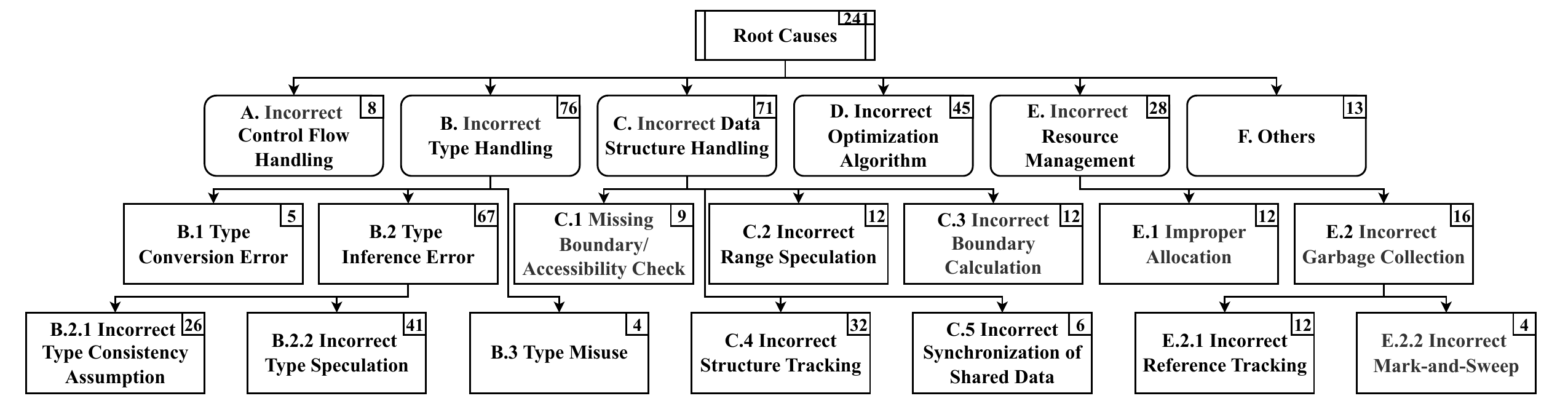}
\captionsetup{justification=centering}
\caption{Root causes of vulnerabilities in JS engine.}
\vspace{-10pt}
\label{fig:rc}
\end{figure*}

% \paragraph{\textbf{Incorrect Control Flow Handling (A)}}
\noindent\textbf{\hypertarget{A}{Incorrect Control Flow Handling (A).}}
These vulnerabilities are primarily caused by the mismanagement of execution flow within JS engines, which includes incorrect code log branching or jumping and inconsistencies between the execution flow logic of two compilations, taking up 3.32\% of the studied vulnerabilities.

% \paragraph{\textbf{Incorrect Type Handling (B)}}
% \noindent\textbf{Incorrect Type Handling (B).}\label{Incorrect Type Handling}
\noindent\textbf{\hypertarget{B}{Incorrect Type Handling (B).}}
Incorrect Type Handling is the most prevalent root cause (31.54\%) among all the vulnerabilities. Given the dynamic type nature of JS, effectively managing variables of different types is crucial. A detailed example is provided in \autoref{appendix:CVE-2019-13728}, which demonstrates how improper assumptions in V8’s handling of \textit{RegExp} objects can lead to OOB.
We further subdivide these vulnerabilities into three subcategories based on the various scenarios in which JS engines process types: \ding{182} \textbf{Type Conversion Error (B.1)} refers to the problems in converting between different internal representations. \ding{183} \textbf{Type Inference Error (B.2)} involves inaccuracies in deducing the types of variables or objects, which includes two subtypes: \textbf{Incorrect Type Consistency Assumption (B.2.1)} vulnerabilities erroneously assumes that a variable’s type remains consistent throughout execution, and \textbf{Incorrect Type Speculation (B.2.2)} vulnerabilities incorrectly speculates the type of variables during the optimization phase. 
% \fish{We observe that this category represents the largest proportion of our studied vulnerabilities, accounting for XXX\% of the total.}
% \ding{184} \textbf{Type Misuse (B.3)} arise when JS engines incorrectly assign a type to objects in a manner that is inconsistent with their expected usage.
\ding{184} \textbf{Type Misuse (B.3)} arises when a type is incorrectly assigned to objects in a manner inconsistent with their intended usage.

\noindent\textbf{\hypertarget{C}{Incorrect Data Structure Handling (C).}}
Incorrect Data Structure Handling accounts for 29.46\% of the studied vulnerabilities. These issues typically stem from improper management or misinterpretation of the JS engine's internal data structures. We have divided these vulnerabilities into five subcategories: \ding{182} \textbf{Missing Boundary/Accessibility Check (C.1)} refers to the missing of necessary checks on the boundaries or accessibility of data structures in JS engines. \ding{183} \textbf{Incorrect Range Speculation (C.2)} involves errors in predicting or speculating the range of valid variables during JS engine optimization. \ding{184} \textbf{Incorrect Boundary Calculation (C.3)} refers to errors where boundary checks exist but are improperly calculated for the specific context. \ding{185}\textbf{Incorrect Structure Tracking (C.4)} covers vulnerabilities caused by failures in maintaining accurate internal metadata or object layouts. \ding{186} \textbf{Improper Synchronization of Shared Data (C.5)} represent the concurrency issues where shared data structures in JS engines are accessed by multiple threads without adequate synchronization, leading to race conditions or data corruption. 
\noindent\textbf{\hypertarget{D}{Incorrect Optimization Algorithm (D).}}
Optimization is a crucial phase in JS engines as it impacts the entire execution process, and associated vulnerabilities can have significant effects. 
% In our study, vulnerabilities due to incorrect optimization algorithms account for \fish{XXX\%} of the analyzed vulnerabilities.
% Incorrect Optimization Algorithm is the third most common root cause among JS engine vulnerabilities, accounting for xxx\%. 
% Optimization is a critical component of modern JS engines. 

% JS engines implement a wide array of optimizations, which we thoroughly investigate and discuss in detail on our website~\cite{website}.

% , ranging from high-level inter-procedural transformations to low-level, machine-specific code generation. 
We find these vulnerabilities often arise from logical errors in the implementation of optimization algorithms within JS engines, which takes up 18.67\% of the vulnerabilities.
% As discussed in ~\autoref{sec:background}, a JS engine implement a wide array of optimizations—from high-level, inter-procedural transformations to low-level, machine-specific code generation. This category involves vulnerabilities that arise from logical errors in the implementation of optimization algorithms in JS engines.
% Figure~\ref{fig:CVE-2017-0234} illustrates CVE-2017-0234~\cite{CVE-2017-0234}, which is a typical vulnerability caused by an incorrect optimization algorithm. This issue occurs when CH's JIT compiler improperly removes boundary checks for virtual typed arrays, assuming all array accesses stay within a 4GB limit. OOB writes happen if this limit is exceeded. The fix introduces additional range checks to prevent unnecessary boundary check removals
A detailed example is provided in \autoref{appendix:CVE-2017-0234}, which illustrates CVE-2017-0234~\cite{CVE-2017-0234}, an optimization-related vulnerability in CH.

% \begin{figure}[htbp]
% \centering
% \includegraphics[width=0.85\linewidth]{img/CVE-2017-0234.pdf}
% \captionsetup{justification=centering}
% \caption{CVE-2017-0234: \textit{Incorrect Optimization Algorithm} vulnerability in CH}
% % \caption{A vulnerability in ChakraCore caused by \textit{Incorrect Optimization Algorithm}\#CVE-2017-0234}
% \label{fig:CVE-2017-0234}
% \end{figure}
% As shown in Figure~\ref{fig:CVE-2017-0234}, which is an example of an \textit{Incorrect Optimization Algorithm} vulnerability, CVE-2017-0234~\cite{CVE-2017-0234}, in which CH’s JIT compiler prematurely eliminates boundary checks for virtual typed arrays. The engine assumes that array accesses will remain within a 4GB limit without confirming the actual index range. Consequently, out-of-bounds writes occur if the index exceeds this limit. The fix adds extra range checks, ensuring that the JIT does not remove necessary bounds checks unless array operations are guaranteed to remain within 4GB. 

% Detailed analysis indicates that these vulnerabilities predominantly arise from flaws in the optimization phase of JIT compilation, where aggressive techniques—such as speculative code motion, dead code elimination, and bounds check elimination—are employed to enhance performance. When these optimization algorithms make incorrect assumptions or misapply transformations, they may inadvertently remove essential runtime checks, miscalculate jump targets, or merge control flow paths improperly. 

% \paragraph{\textbf{Incorrect Resource Management (E)}}
\noindent\textbf{\hypertarget{E}{Incorrect Resource Management (E).}}
These vulnerabilities, accounting for 11.62\% of the total, result from the mismanagement of various resources within the JS engine, which includes three subcategories: 
% \ding{182} \textbf{Improper Allocation (E.1)} encompasses errors related to the allocation and initialization of resources, including memory, objects, and other execution-related entities. It covers incorrect memory allocation, improper initialization of variables or objects, and failures in de-allocation. 
\ding{182} \textbf{Improper Allocation (E.1)} involves errors in allocating or initializing resources such as memory and objects, including incorrect allocation, improper initialization, and deallocation failures.
\ding{183} \noindent\textbf{\hypertarget{E.2}{Incorrect Garbage Collection (GC) (E.2)}}
% \textbf{Incorrect Garbage Collection (GC) (E.2)}
involves errors in the mechanisms responsible for reclaiming unused resources. These vulnerabilities can be divide into \textbf{Incorrect Reference Tracking (E.2.1)}, which pertains to errors in maintaining correct reference counts or relationships between resources and \textbf{Incorrect Mark-and-Sweep (E.2.2)}, which refers to flaws in the process of marking active resources and subsequently sweeping unreferenced ones. 
% 3) \textbf{Incorrect Register Management (E.3)}: this subcategory refers to vulnerabilities arising from the mismanagement of CPU or virtual registers. It encompasses errors in the allocation, usage, and release of registers during code execution, affecting the handling of temporary data and computational state. 

% \paragraph{\textbf{Others (F)}}
\noindent\textbf{\hypertarget{F}{Others (F).}}
For vulnerabilities that fail to fit into our established categories, we assemble them in the category titled \textit{Others}, such as vulnerabilities caused by incorrect handling of register and wrong resolution for the engine parameter scope~\cite{CVE-2017-2531}.

\noindent
\begin{tcolorbox}[colback=gray!15, colframe=black,boxrule=1pt,before=\vspace{1pt},after={\vspace{0pt}},left=1pt,right=1pt,top=0.5pt,bottom=0.5pt]
\textit{\textbf{Finding 3}: \label{Finding 3}
% \nameref{Incorrect Type Handling} and \nameref{Incorrect Data Structure Handling} 
\textbf{\hyperlink{B}{Incorrect Type Handling (B)}} and \textbf{\hyperlink{C}{Incorrect Data Structure Handling (C)}} account for a majority of studied vulnerabilities, reflecting the challenges posed by JS’s dynamic type system and flexible object structures.}
\end{tcolorbox}

% A more detailed breakdown of how these root causes manifest across different JS engines is provided in Appendix \ref{appendix:rootcause-distribution}. The analysis demonstrates that each engine exhibits unique vulnerability patterns aligned with its internal architecture. Specifically, we observe that ChakraCore is dominated by type-handling errors, V8 and JSC by data-structure handling issues, and SpiderMonkey by GC-related faults.

% \begin{figure}[t]
% \centering
% \includegraphics[width=0.98\linewidth]{img/CVE-2018-16065.pdf}
% \captionsetup{justification=centering}
% \caption{\fishdel{CVE-2018-16065: \textit{Incorrect Data Structure Handling} vulnerability in V8}}
% \label{fig:CVE-2018-16065}
% \end{figure}

\begin{figure}[t]
\centering
\includegraphics[width=0.9\linewidth]{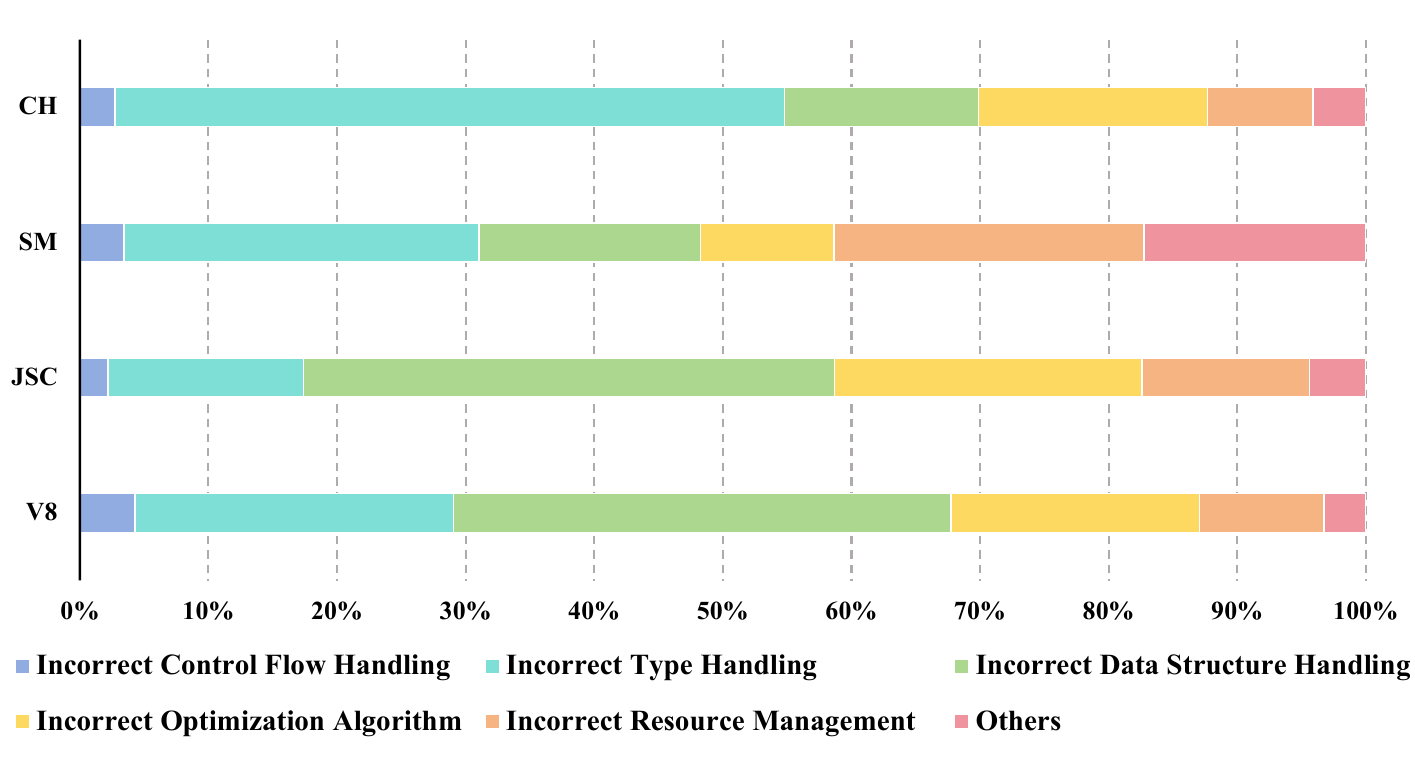}
\captionsetup{justification=centering}
\caption{Vulnerability distribution in each JS engine by Root Causes}
% \vspace{-2em}
\label{fig:rtd}
\end{figure}

\subsection{Root Cause Distribution Analysis}
\autoref{fig:rtd} illustrates the distribution of vulnerabilities in four JS engines, categorized by their root causes. We observe that vulnerabilities related to JS property management 
% (including \nameref{Incorrect Type Handling} and \nameref{Incorrect Data Structure Handling}) 
(including \textbf{\hyperlink{B}{Incorrect Type Handling (B)}} and \textbf{\hyperlink{C}{Incorrect Data Structure Handling (C)}}) 
constitute the majority in all engines. 

% V8 (63.4\%), JSC (56.5\%), CH (67.1\%), and SM (44.8\%). 

Notably, CH exhibits a significantly higher proportion of 
% \nameref{Incorrect Type Handling}
\textbf{\hyperlink{B}{Incorrect Type Handling (B)}}
vulnerabilities (52.05\%) compared to the other engines. This discrepancy stems from CH’s design, which is less sensitive to dynamic type conversions and structural changes. Such insensitivity often leads to inconsistencies, compounded by overly aggressive speculative optimizations that lack adequate fallback mechanisms.
In contrast, V8 and JSC show considerably higher proportions of 
% \nameref{Incorrect Data Structure Handling}
\textbf{\hyperlink{C}{Incorrect Data Structure Handling (C)}} vulnerabilities (38.71\% and 41.30\%, respectively). 
In JSC, these issues mainly stem from failures to maintain structural integrity during dynamic property updates, prototype transitions, and array operations. 
In V8, the issue arises from the complexity of its inline caching and hidden class system, which is designed to enhance execution speed~\cite{v8hiddenclasses} but also introduces risks of type instability and structural inconsistencies. Moreover, V8’s Typer phase demonstrates challenges with speculative optimization and integer conversion in edge cases.

We also observe that SM exhibits a lower proportion of 
% \nameref{Incorrect Optimization Algorithm}
\textbf{\hyperlink{D}{Incorrect Optimization Algorithm (D)}}
vulnerabilities (10.34\%) compared to other engines. 
Our analysis indicates that vulnerabilities in SM primarily stem from instruction reordering and function inlining rather than aggressive speculative optimizations like TCE or DCE, which are more prevalent in V8. This may be attributed to SM’s more conservative optimization strategy, which emphasizes correctness over speculative performance gains.
Furthermore, SM exhibits the highest proportion of \textbf{\hyperlink{E.2}{Incorrect GC (E.2)}} vulnerabilities (24.14\%), far exceeding other JS engines. This suggests that its GC design faces challenges in reference tracking and object lifetime management, leading to frequent UAF issues.

\noindent
\begin{tcolorbox}[colback=gray!15, colframe=black,boxrule=1pt,before=\vspace{1pt},after={\vspace{0pt}},left=1pt,right=1pt,top=0.5pt,bottom=0.5pt]
\textit{\textbf{Finding 4}: \label{Finding 4}
Different JS engines exhibit predominant vulnerability types driven by their design mechanisms.
% , such as type handling, optimization, and GC.
CH is more prone to type handling errors, while V8 and JSC frequently have vulnerabilities when handling data structures. 
% SM employs a conservative optimization approach and 
SM primarily exhibits a higher rate of GC-related issues.}
% \fei{the comparisons are divergent and hard to get the main points. Need a finding to summarize the different implementation focuses of the engines and their impact on the vuls distributions.}
\end{tcolorbox}
% \textbf{Commonality.} 
% As shown in Figure~\ref{fig:c}, by analyzing the root causes of vulnerabilities across the four major JS engine and calculating their Spearman correlation coefficients for these root cause distributions, we observe the following: V8 and JavaScriptCore show the highest correlation (0.639), followed by ChakraCore and JavaScriptCore (0.564) and SpiderMonkey and JavaScriptCore (0.535), whereas SpiderMonkey and ChakraCore exhibit the lowest correlation (0.318). Overall, the correlations range from about 0.3 to the mid-0.6 level, suggesting a moderate degree of similarity without being extremely uniform or entirely divergent. 
% In summary, the moderate correlation suggests that while these engines share certain broad characteristics—such as having to manage dynamic types, object structures, concurrency, and security—they still retain distinct design choices and optimization tactics leading to marked differences in bug root causes. 

% \textbf{Finding 7}:  A moderate correlation (0.3–0.6) in vulnerability root causes suggests that while JS engines share common challenges, distinct design choices lead to noticeable differences. 
% \section{Attack and Defense}
\section{Vulnerability Exploitation} \label{sec:vulnerabilitytrigger}

% \begin{figure*}[t]
% \centering
% \includegraphics[width=0.94\linewidth]{img/chain.pdf}
% \captionsetup{justification=centering}
% \caption{Vulnerability Trigger Chain of vulnerabilities in JS engine.}
% \label{fig:rc}
% \end{figure*}

The severity of vulnerabilities largely stems from their exploitability by attackers, which can lead to significant security risks. After understanding the symptoms and root causes of JS vulnerabilities, we delve into how these vulnerabilities can be exploited, offering insights into their exploitability and security implications.

Among various exploitation strategies, \textbf{MSVs} play a particularly critical role, as they often serve as the foundation for severe attacks such as arbitrary code execution and sandbox escapes. 
% While not all vulnerabilities immediately result in MSVs, many logical flaws propagate through JS engine phases, ultimately leading to MSVs that attackers can exploit.
% To systematically examine this escalation process, we conduct an exploratory study on vulnerability trigger chains, which trace how initial errors propagate across different phases, providing insights into the transformation from logical errors to exploitable MSVs.
Although not every vulnerability immediately produces an MSV, many logical flaws propagate through multiple JS-engine phases and eventually culminate in exploitable MSVs. To study this escalation, we perform an exploratory analysis of vulnerability trigger chains that trace how initial logical errors traverse engine phases and evolve into MSVs, yielding insights for detection and mitigation.

\subsection{Vulnerability Trigger Chain Construction}
Given a vulnerability, our goal is to identify its trigger chain that can ultimately lead to MSVs. It is important to note that the vulnerability may not directly cause such violations but can affect subsequent JIT optimization or execution phases, leading to a cascade of errors that eventually result in MSVs, as detailed in JS engine phases in \autoref{sec:background}.
% for the introduction of the phases of the JS engine.
Formally, we define the vulnerability trigger chain as:

\begin{definition} [Vulnerability Trigger Chain] \label{def-mr}  
The trigger chain of a vulnerability \( V \), which originates in phase \( P_0 \) with root cause \( R \), is defined as a sequence \(\langle R, P_0, S_0, P_1, S_1, \ldots, P_n, S_n \rangle\), where:

\begin{itemize}[label={--},leftmargin=*]
    \item \( R \) represents the root cause of the vulnerability. 
    \item   \( P_0 \) denotes the initial execution phase within the JS engine where the vulnerability first manifests.
    \item \( P_i \) (\( i > 0 \)) represents the phase affected by the preceding symptom \( S_{i-1} \).
    \item \( S_i \) is the symptom observed after the execution of the current phase \( P_i \).  
    \item The final node, \( S_n \), is a MSV, where the chain culminates. 
\end{itemize}
\end{definition}

Intuitively, a vulnerability trigger chain represents the propagation of errors across different phases (introduced in \autoref{sec:background}) of a JS engine (from \( P_0 \) to \( P_n \)). Each symptom \( S_i \) (summarized in \autoref{sec:symptom}) is a consequence of the error propagation within a specific phase (\( P_i \)). Additionally, the root cause (defined in \autoref{sec:rootcause}) of the vulnerability (\( R \)) is included at the beginning of the chain, preceding the first phase (\( P_0 \)) where the vulnerability manifests.

% A vulnerability trigger chain $R \to P \to S$ contains three components, the vulnerability's root cause~(defined in Figure~\ref{sec:rootcause}) $R$, its affected work phase in JS engine~(introduced in Section~\ref{sec:background}) $P$, and the final symptom~(summarized in Section~\ref{sec:symptom}) $S$. 

To build and verify comprehensive vulnerability trigger chains that cover as many collected vulnerabilities as possible, we design a chain construction paradigm as follows:

\begin{itemize}[leftmargin=*]
\item To systematically analyze vulnerability trigger chains, 
we select vulnerabilities with known exploit scripts, as they offer concrete runtime data that helps us model the underlying trigger chains accurately.
Since not all MSVs are directly exploitable, we focus on vulnerabilities that have been successfully exploited in real-world attacks.  
Selecting exploitable vulnerabilities ensures that our study focuses on security issues that pose actual threats, rather than those that merely cause program crashes without affecting control flow. 
Moreover, known exploits often come with detailed reports or PoCs that illustrate error propagation, enabling accurate construction and validation of trigger chains.
% Additionally, vulnerabilities with known exploits often come with detailed research, disclosure reports, or PoC, providing clear insights into error propagation. These concrete examples help us construct and validate vulnerability trigger chains, making our analysis more robust and credible.  
% Our analysis reveals that 75 out of 241 vulnerabilities in our dataset contain relevant exploit information, offering concrete examples of how MSVs can be exploited. 
Our study identifies 75 out of 241 vulnerabilities with such exploit data, enabling credible and detailed chain analysis.

% \item For each identified vulnerability, we construct vulnerability trigger chains by manually analyzing how it propagates through different JS engine phases before leading to MSVs. 
% Since we have already identified the root cause and symptom of each vulnerability, we begin by executing its PoC to validate its behavior and confirm its initial impact.
% We then track how the error propagates through different phases, recording each phase it affects and the corresponding symptom at that stage.
% Our analysis indicates that 43 out of the 75 vulnerabilities exhibit a multi-step propagation process, where intermediate errors contribute to the escalation before leading to MSVs. Meanwhile, the remaining 32 vulnerabilities directly result in MSVs without any intermediate propagation stages. 
\item For each identified vulnerability, we manually construct its trigger chain by analyzing how it propagates across JS engine phases before resulting in MSVs. Starting from the PoC, we execute the exploit to observe its runtime behavior and confirm the initial error manifestation. We then trace how the error propagates through subsequent engine phases, documenting the specific symptoms at each stage. Our analysis shows that 43 of 75 vulnerabilities involve multi-step propagation, while the remaining 32 directly lead to MSVs without intermediate stages.

% \item Finally, we systematically organize the extracted information to construct the complete vulnerability trigger chain as defined in Definition~\ref{def-mr}. This involves structuring the identified root cause, affected phases, and observed symptoms into a coherent sequence, capturing how the vulnerability propagates through the JS engine. These structured trigger chains allow us to analyze common propagation patterns and escalation mechanisms across different vulnerabilities.
\item Finally, we organize the extracted information into complete vulnerability trigger chains (\autoref{def-mr}) by structuring the root cause, affected phases, and corresponding symptoms into coherent sequences. These chains capture how vulnerabilities propagate through the JS engine, enabling analysis of common escalation patterns across different cases.

\end{itemize}

\begin{figure*}[t]
    \centering
    \begin{subfigure}[b]{0.75\linewidth}
        \centering
        \includegraphics[width=\linewidth]{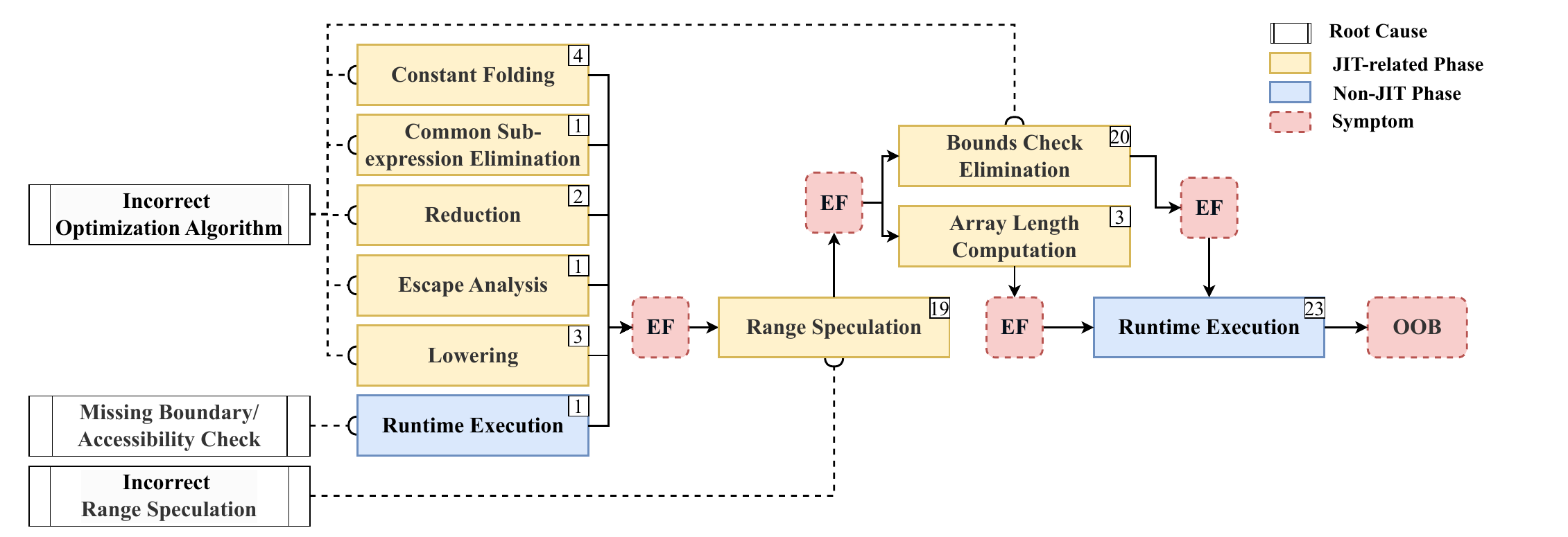}
        \caption{Vulnerability trigger chains leading to OOB}
        \label{fig:sub1}
    \end{subfigure}

    \begin{subfigure}[b]{0.75\linewidth}
        \centering
        \includegraphics[width=\linewidth]{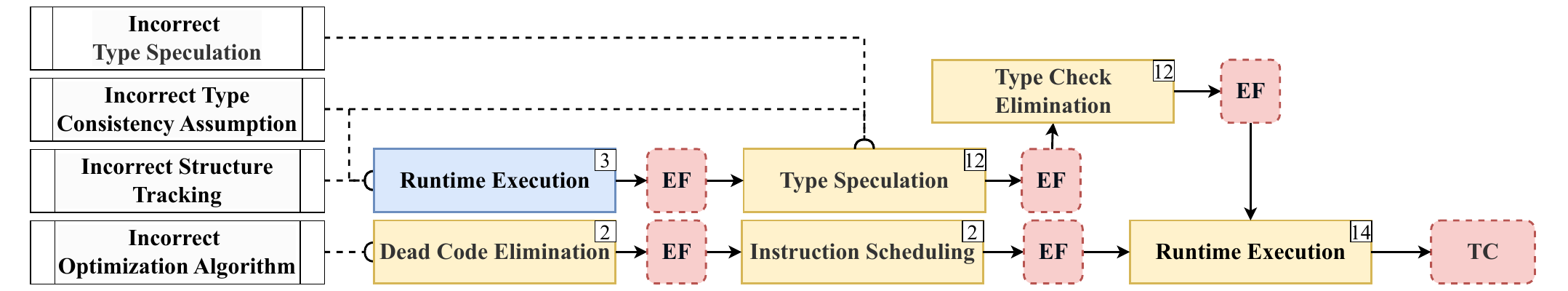}
        \caption{Vulnerability trigger chains leading to TC}
        \label{fig:sub2}
    \end{subfigure}
    \begin{subfigure}[b]{0.75\linewidth}
        \centering
        \includegraphics[width=\linewidth]{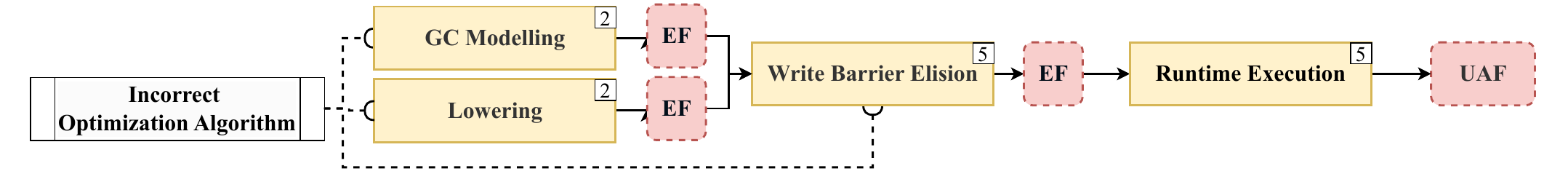}
        \caption{Vulnerability trigger chains leading to UAF}
        \label{fig:sub3}
    \end{subfigure}
    \captionsetup{justification=centering}
    \caption{Vulnerability trigger chains in JS engines.}
    \label{fig:tc}
\end{figure*}

In total, we identified 75 vulnerability trigger chains, including 32 single-phase cases that directly lead to memory violations, which have been covered in \autoref{sec:rootcause}. Our analysis here focuses on the remaining complex 43 multi-phase chains, which exhibit 22 distinct propagation patterns. To facilitate understanding, we group these multi-phase chains into three graph structures according to their final MSV types, including OOB, TC, and UAF, as shown in \autoref{fig:tc}. Each graph visualizes unique propagation patterns converging on the same MSV type. A directed path denotes one such pattern, tracing the flow from the root cause to the resulting MSV, while numeric labels on phase nodes indicate the number of vulnerabilities (out of 43) that originate from or traverse that phase.

% Finally, we identified 75 vulnerability trigger chains, of which 32 are single-phase chains where the vulnerability immediately results in a memory violation without further propagation. To focus on more complex cases, we primarily analyzed the remaining 43 multi-phase chains, which correspond to 22 unique propagation patterns.

% To simplify representation, we group vulnerability trigger chains that culminate in the same type of MSV (i.e., OOB, TC, and UAF) into a graph structure.  
% Figure~\ref{fig:tc} illustrates the categorized vulnerability trigger chains in three distinct graphs, each representing one type of MSV.  
% The number associated with each phase represents the total count of vulnerabilities (out of the 43 identified) that either originate in this phase or pass through it during propagation.

% To simplify representation, we group the 43 multi-phase vulnerability trigger chains into three graph structures based on the final type of MSV they produce—namely, OOB, TC, and UAF.  
% Each graph in \autoref{fig:tc} visualizes multiple unique propagation patterns leading to the same type of MSV.  
% In total, we identify 22 distinct trigger chain patterns, where each pattern corresponds to a unique sequence of engine phases through which errors propagate.  
% Within each graph, a directed path represents one such pattern, tracing the flow from a specific root cause to the final MSV.  
% The numeric labels attached to each phase node indicate the total number of vulnerabilities (out of the 43) that either originate or pass through that phase during their propagation.

\subsection{Exploitation Analysis}

\autoref{fig:sub1} illustrates how vulnerabilities initially manifesting as EF vulnerabilities can escalate into OOB. 
% We observe that 86.96\% of OOB vulnerabilities in Figure~\ref{fig:sub1} arise from the incorrect removal of essential bounds checks. However, these vulnerabilities do not all originate in the BCE phase itself but stem from errors in earlier optimization or execution phases, which propagate step by step and ultimately lead to BCE-related errors.
Although 86.96\% of OOB vulnerabilities in \autoref{fig:sub1} involve incorrect bounds check removal, they often originate in earlier phases, with errors propagating and ultimately causing BCE-related issues.
For instance, incorrect numerical computations during CF or Escape Analysis, as well as missing boundary or accessibility checks at runtime, can introduce Range Speculation errors. Such errors may prompt elimination of necessary bounds checks and ultimately lead to OOB.

For example, CVE-2020-9802~\cite{CVE-2020-9802} originates from an incorrect optimization algorithm during the CSE phase. Through the attacker's careful crafting, the incorrect CSE implementation triggers a cascade of optimization errors, starting with incorrect numeric results. Since range speculation determines the potential value ranges of variables based on accurate expression values, the erroneous CSE outputs lead to range speculation being based on incorrect information. 
As a result, the BCE phase wrongly eliminates necessary bounds checks, ultimately causing OOB access at runtime.
% Subsequently, the BCE phase improperly removes array access boundary checks that should have been retained. This chain of errors ultimately leads to OOB access at runtime.

% Figure~\ref{fig:sub2} illustrates how EF propagates through multiple phases, ultimately leading to TC. 
\autoref{fig:sub2} shows how EF propagates across phases to cause TC.
Errors in Runtime Execution propagate to Type Speculation, leading to incorrect inferences that cause the erroneous removal of type checks in the TCE phase. This weakens type safety, allowing unintended type assumptions to persist into Runtime Execution. 
Separately, DCE may remove essential operations, disrupting Instruction Scheduling and impacting Runtime Execution.
% For instance, this could sever a dependency on an operation that triggers deoptimization, thereby causing TC.
% \fei{what do you mean for "sever a dependency"}

\autoref{fig:sub3} illustrates how incorrect optimization algorithms in the Lowering, GC Modeling, and WBE phases can lead to UAF.
Errors propagate into WBE, causing the engine to skip write barriers. This leads to premature object reclamation during Runtime Execution, resulting in UAF when the engine attempts to access a freed object.

% Among these chains, 90.70\% of vulnerabilities are linked to the JIT compiler, mainly due to faulty optimizations or speculation. The remaining 9.30\% originate from runtime or interpreter issues, which can also be exploited via JIT interactions. This highlights JIT compilation as a major attack surface: speculative execution, dynamic code generation, and aggressive optimizations can inadvertently introduce security flaws. Since each JIT phase depends on the accuracy of previous phases, early misjudgments—such as incorrect type or control flow assumptions—can cascade unchecked, escalating into exploitable conditions.
Among these chains, 90.70\% of vulnerabilities are linked to the JIT compiler, mainly due to faulty optimizations or speculation, while the remaining 9.30\% stem from runtime or interpreter issues exploitable through JIT interactions. This underscores JIT compilation as a major attack surface, where speculative execution and aggressive optimizations can introduce security flaws. Because each JIT phase depends on prior analyses, early misjudgments, such as incorrect type or control flow assumptions, can cascade into exploitable conditions.

\noindent
\begin{tcolorbox}[colback=gray!15, colframe=black,boxrule=1pt,before=\vspace{1pt},after={\vspace{0pt}},left=1pt,right=1pt,top=0.5pt,bottom=0.5pt]
\textit{\textbf{Finding 5}\label{Finding 5}: 
% JIT compilation is the main enabler of vulnerability propagation, with most exploits leveraging its optimizations directly or indirectly. Speculative assumptions in JIT phases can amplify minor logical errors into exploitable vulnerabilities, making JIT a critical attack surface.
JIT compilation plays a key role in vulnerability propagation, as most exploits leverage its optimizations. Speculative assumptions in JIT phases can amplify minor logic flaws into exploitable vulnerabilities, making JIT a critical attack surface.
}
\end{tcolorbox}

From the constructed vulnerability trigger chains, we also find that key optimization phases (e.g, BCE, TCE, WBE) substantially amplify exploitability by transforming minor logic flaws into severe security violations. The dominant pattern, observed in 69.77\% of the 43 analyzed cases shown in \autoref{fig:tc}, is \textbf{speculative safety mechanism removal}, where incorrect assumptions about program state lead to the premature elimination of essential checks. Specifically, 37.21\% of cases arise from range speculation errors causing BCE to remove memory-safety checks (triggering OOB issues), 27.91\% stem from type speculation errors that prompt TCE to skip type verification (leading to type confusion), and 4.65\% result from GC modeling errors causing WBE to omit memory protection (resulting in UAF). These optimization phases thus serve as critical escalation points in the exploitation chain.

\noindent
\begin{tcolorbox}[colback=gray!15, colframe=black,boxrule=1pt,before=\vspace{1pt},after={\vspace{-3pt}},left=1pt,right=1pt,top=0.5pt,bottom=0.5pt]
\textit{\textbf{Finding 6}\label{Finding 6}: Speculative safety mechanism removal serves as a key mechanism for transforming minor logical errors into exploitable MSVs, particularly in BCE, TCE, and WBE.}
\end{tcolorbox}

\begin{figure}[t]
\centering
\includegraphics[width=0.9\linewidth]{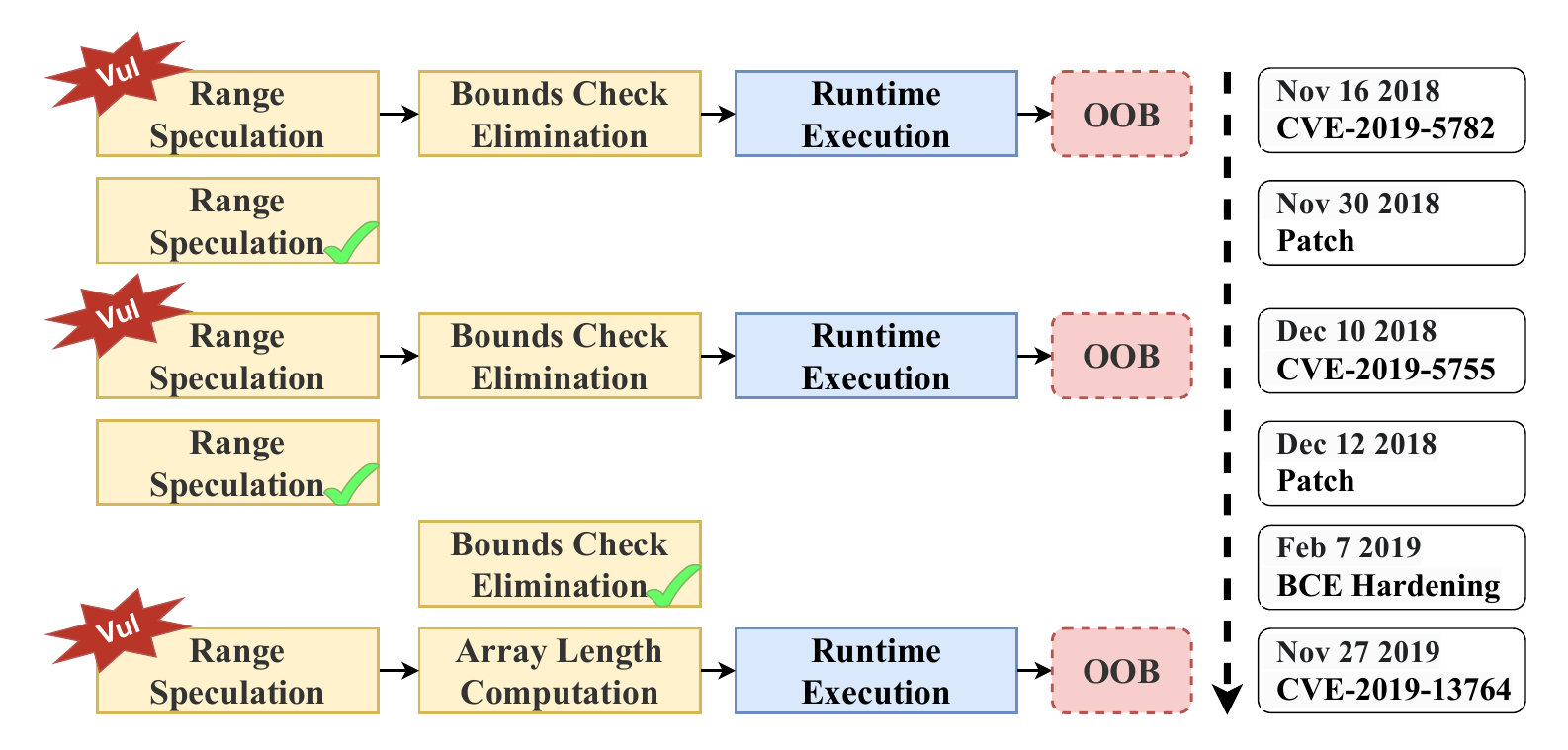}
\captionsetup{justification=centering}
\caption{Examples of exploitation and mitigation evolution}
\vspace{-15pt}
\label{fig:timeline}
\end{figure}

\noindent
\subsection{Mitigation Analysis} 
% To defend against potential attacks, it is essential to disrupt the vulnerability trigger chain to prevent memory errors from occurring. 
% To defend against potential attacks, disrupting the trigger chain is essential.
% A straightforward approach is to repair vulnerabilities directly based on their root causes. 
% Our analysis shows that all 43 vulnerabilities in Figure~\ref{fig:tc} have been fixed based on their root causes.
% However, while this may fix specific vulnerabilities, the vulnerability trigger chain remains intact, meaning that new vulnerabilities may still emerge in the future, exploiting the same exploitation techniques.  
To defend against potential attacks, it is crucial to disrupt the vulnerability trigger chain. A direct approach is to patch vulnerabilities at their root causes, and our analysis shows that all 43 cases in \autoref{fig:tc} were fixed in this way. However, although such fixes address specific issues, the trigger chains often persist, allowing similar vulnerabilities to reappear through the same exploitation techniques.

\autoref{fig:timeline} illustrates the evolution of Range Speculation-based exploits in V8 and the corresponding mitigation strategies. The earliest vulnerability, CVE-2019-5782~\cite{CVE-2019-5782}, exploited Range Speculation to mislead BCE, ultimately leading to OOB access during runtime execution. This vulnerability was initially mitigated through a patch (Nov 30, 2018) that strengthened Range Speculation handling. However, while that CVE was fixed, similar vulnerabilities (e.g., CVE-2019-5755~\cite{CVE-2019-5755}) continued to emerge, leveraging the same attack trigger chain. 
Recognizing that individual patches were insufficient to fully prevent such exploits, developers introduced a broader security mechanism, BCE Hardening~\cite{BCEHardening} (Feb 7, 2019), to structurally mitigate vulnerabilities that can affect BCE. 
This enhancement can block the previous vulnerability trigger chain, which significantly reduces the feasibility of BCE-based exploitation. 
% However, while this mitigates a specific attack vector, other speculative optimizations may remain exploitable.
However, other speculative optimizations may still be exploitable.
For example, attackers adapted their strategies, shifting from BCE manipulation to exploiting Range Speculation in conjunction with Array Length Computation, as demonstrated in CVE-2019-13764~\cite{CVE-2019-13764} (Nov 27, 2019). This bypass technique allowed them to regain OOB access through another trigger chain.

% The progression of these exploits highlights that while targeted patches can mitigate individual vulnerabilities, comprehensive hardening of critical phases (e.g., BCE, TCE, and WBE) is essential for disrupting entire vulnerability trigger chains and significantly reducing practical exploitability. Additionally, identifying as many distinct trigger chains as possible is crucial, as it further minimizes attack opportunities.
These progression of these exploits show that while targeted patches address individual issues, hardening critical phases (e.g., BCE) and identifying diverse trigger chains are key to disrupting exploitation paths and reducing attack surfaces.

% It underscores the adaptive nature of attack development, as attackers continuously evolve their techniques to bypass mitigations. 

\noindent
\begin{tcolorbox}[colback=gray!15, colframe=black,boxrule=1pt,before=\vspace{1pt},after={\vspace{0pt}},left=1pt,right=1pt,top=0.5pt,bottom=0.5pt]
\textit{\textbf{Finding 7}\label{Finding 7}: Fixing individual vulnerabilities is insufficient to prevent attacks, whereas blocking the trigger chain by hardening critical phases can effectively mitigate similar exploits. However, attackers may adapt by identifying alternative chains to bypass defenses and launch attacks.}
\end{tcolorbox}

\section{Discussion}
% \subsection{Implications}
% \fish{This part can be summarized.}
Through our in-depth analysis of vulnerability characteristics and their exploitation in JS engines, we summarize key implications based on our analysis, particularly for developers and researchers.

% \noindent
% \textit{\textbf{7.1.1 For Developers.}}
\subsection{For Developers}

\noindent\ding{182} \textbf{Be Careful in Handling Types and Data Structures.}  
As highlighted in \textbf{Finding~3}, a large proportion of JS engine vulnerabilities arise from improper handling of dynamic types and internal structures. Developers should therefore enforce strict safeguards—such as validating type conversions and property updates (e.g., array length or object layout changes) to prevent unintended memory layout shifts, and periodically verifying speculative optimizations to ensure consistency between assumed and actual runtime types. Moreover, as noted in \textbf{Finding~4}, security measures should be tailored to engine-specific weaknesses: ChakraCore is more prone to type conversion errors, while V8 and JSC frequently exhibit data-structure inconsistencies. Applying these checks during property addition, array expansion, and prototype modification can substantially reduce data handling vulnerabilities.

% \item 
% \noindent\ding{182} \textbf{Be Careful in Handling Types and Data Structures.} 
% As \textbf{Finding 3} shows, many JS engine vulnerabilities stem from mishandling dynamic types and structures.
% To mitigate these risks, developers should focus on several key safeguards: strictly controlling type conversions and property updates (e.g., verifying array length or object property changes to prevent unintended memory layout alterations), periodically validating speculative optimizations (e.g., checking whether the engine’s assumed “type information” matches actual runtime conditions, and rolling back if discrepancies are found), and tailoring security measures to each engine’s unique weaknesses (as \textbf{Finding 4} notes, CH often faces errors during type conversion, while V8 and JSC tend to have inconsistencies in data structures). 
% Applying these safeguards at key points such as adding properties, expanding arrays, or modifying prototypes can greatly reduce JS data handling risks.

\noindent\ding{183} \textbf{Balance JS Engine Efficiency and Security during Optimization.}
As highlighted in \textbf{Finding~5}, the optimization pipeline substantially broadens the attack surface, making security a first-class concern. \textbf{Finding~6} further shows that speculative safety mechanism removal (e.g., BCE, TCE, and WBE) can escalate minor logic errors into severe memory-safety violations, while \textbf{Finding~4} indicates that conservative optimization strategies may mitigate such risks. Yet, these aggressive optimizations remain indispensable for performance. To reconcile efficiency and security, developers should integrate security-aware mechanisms into the optimization pipeline. 
\noindent\ding{184} \textbf{Fix Vulnerabilities by Considering the Entire Exploit Chain.} As highlighted in \textbf{Finding 7}, the recurring cycle of exploit and mitigation suggests that fixing individual vulnerabilities in isolation may be insufficient against evolving attack strategies. 
We recommend that mitigation efforts take a holistic view of the exploit chain and consider how errors in earlier phases propagate into security-critical failures. For instance, OOB vulnerabilities often stem from issues in BCE or incorrect array length speculation. 
Therefore, in addition to patching, developers may consider reinforcing critical optimization phases (e.g., via hardening~\cite{BCEHardening}) and enriching regression test coverage to better detect such propagation patterns~\cite{V8RegressCVE-2019-5755,V8RegressCVE-2019-5782}.

\subsection{For Researchers}

\noindent\ding{182} \textbf{Detect Vulnerability Trigger Chains.}
As highlighted in \textbf{Finding~6}, analyzing vulnerability trigger chains is crucial for uncovering exposed attack surfaces and identifying potential patch bypass paths, particularly in memory-safety violations. Such analysis enables more targeted testing and mitigation strategies. However, efficiently discovering new trigger chains remains challenging due to the high cost of manual analysis. Recent advances in LLMs and agentic AI offer a promising alternative: LLM-based agents can systematically explore vulnerability trigger chains and detect latent weak points~\cite{wei2024llm,guo2025repoaudit}, functioning as automated, continuous security auditors that strengthen JS engine resilience against exploitation.

    \noindent\ding{183} \textbf{Design Bug-Specific Testing Methods.}Our study reveals numerous JS-specific vulnerabilities, underscoring the need for tailored testing strategies beyond general-purpose fuzzing. As highlighted in \textbf{Finding 3}, many vulnerabilities originate from \emph{Incorrect Type Handling} and \emph{Incorrect Data Structure Handling}.
    % Researchers can improve existing fuzzing methods by targeting the speculation mechanisms and optimization strategies of JS engines, given their frequently reported vulnerabilities. The unique characteristics of JS engines, such as \emph{implicit type conversions}, \emph{speculative type inference}, and \emph{dynamic structure transitions}, demonstrate particular fragility, which could guide the design of effective test inputs and feedback mechanisms, providing focused directions for enhancing testing methodologies.
    % Researchers can enhance fuzzing by targeting the speculation and optimization mechanisms of JS engines, which are frequent sources of vulnerabilities. Fragile features such as \emph{implicit type conversions}, \emph{speculative type inference}, and \emph{dynamic structure transitions} provide clear guidance for designing targeted test inputs and feedback strategies.
    Fuzzing effectiveness can be improved by focusing on JS engines’ speculation and optimization mechanisms, which are frequent sources of vulnerabilities. Fragile features like \emph{implicit type conversions}, \emph{speculative type inference}, and \emph{dynamic structure transitions} offer concrete directions for designing targeted inputs and feedback strategies.

\noindent\ding{184} \textbf{Enhance JS Engine Testing with Advanced Oracles.}  
Although prior research has proposed various testing approaches for JS engines, our findings indicate that certain vulnerabilities—such as EF errors, which remain difficult to detect due to the absence of crashes or other observable abnormal behaviors. Moreover, as illustrated in \autoref{fig:tc}, some logical bugs do not manifest as memory violations yet still introduce security risks. A promising direction is to develop advanced testing oracles guided by our vulnerability trigger chains, enabling earlier detection of logical flaws before they escalate into exploitable states. Such oracles could also contribute to vulnerability coverage and effectiveness of JS engine testing.

\section{Related Works}

% We review the related works from the perspectives of empirical studies for compiler bugs and automated testing for JS engines.

\textbf{Empirical Study for Compiler Bugs. }
Existing empirical studies on compiler bugs have provided insights into the common faults and their characteristics in various compilers, they often focus on general-purpose compilers like GCC and LLVM. Specifically, Sun et al.~\cite{sun2016toward} conducted the empirical study on bug features in two mainstream compilers, and Zhou et al~\cite{zhou2021empirical} deepened the study, aiming to understand the optimization bugs within the same domain. 

For prior studies on JavaScript engines, Wang et~al.~\cite{wang2023empirical} examined bugs in three major engines (V8, SM, and CH) but focused on general defect characteristics rather than security vulnerabilities or exploit behavior. The \textit{js-vuln-db} project~\cite{js-vuln-db} collected 178 JS engine CVEs with proofs of concept, yet its coverage is outdated and limited, primarily encompassing pre-2020 data. Kang~\cite{kang2021review} provided a review of JS engine vulnerability research, emphasizing memory-related flaws and testing methodologies. In contrast, our work offers the largest and most up-to-date dataset to date with comprehensive examination of exploitation chains and mitigation strategies.

\noindent
\textbf{Automated Testing for JS Engines.}
Previous research has focused on the automated testing of JS engines, primarily employing fuzz testing methodologies.
% Automated testing JS engines have advanced significantly with the development of fuzzing methodologies, which have gained considerable attention as effective techniques for enhancing the security of JS engines.
Specifically, Jsfunfuzz~\cite{jsfunfuzz} generates test cases based on predefined grammar rules, and Domato~\cite{domato} leverages context-free grammar for testing. % Many works also adopt mutation-based fuzzing techniques for testing JS engines.
LangFuzz~\cite{holler2012langfuzz} generates JS test cases by replacing code fragments of PoCs of historical JS engine vulnerabilities.
Superion~\cite{wang2019superion} extends AFL~\cite{afl} to perform grammar-aware trimming and tree-based mutations on the ASTs. 
% DIE~\cite{park2020fuzzing} advances AST-based mutation by stochastically preserving aspects that are likely to expose bugs across mutation.
SoFi~\cite{he2021sofi} introduces fine-grained analysis and reflection-based augmentation for mutation for more valid and diverse test samples. JIT-picking~\cite{bernhard2022jit} and FuzzJIT~\cite{wang2023fuzzjit} are designed specifically to detect JIT compiler bugs by comparing the differences in the outcomes of JS functions before and after JIT compilation. 
% Fuzzilli~\cite{gross2023fuzzilli} enables finer-grained mutations based on a specially designed Intermediate Representation (IR). 
OptFuzz~\cite{wang2024optfuzz} proposed an optimization-path-guided fuzzing framework designed for JS JIT compilers. 

\section{Conclusion}
In this paper, we present a new dataset containing 241 CVEs related to JS engines across four major engines, along with a systematic analysis of these vulnerabilities. Our analysis provides a robust taxonomy of the symptoms and root causes of the collected vulnerabilities. Furthermore, to enhance the understanding of security risks in JS engines, we explore the exploitation potential of these vulnerabilities by reconstructing vulnerability trigger chains and discussing mitigation approaches.
We further analyze the identified root causes, symptoms, and vulnerability trigger chains, leading to six key findings. Based on these findings, we summarize several critical implications for various stakeholders, including developers and researchers. We believe these insights serve as a clear guide for future research and development efforts toward enhancing the security and reliability of JS engines.

% \clearpage

% \bibliographystyle{unsrt}
\bibliographystyle{ACM-Reference-Format}
\bibliography{www26}

% \clearpage
\appendix
\section{Appendix}
\addcontentsline{toc}{section}{Appendix}

\subsection{Data Collection and Screening}
\label{appendix:datacollection}

We select the most popular JS engine projects~\cite{w3-survey,statcounter-survey}, specifically \textit{V8}, \textit{JSC}, \textit{SM}, and \textit{CH}.

To ensure the comprehensiveness of our dataset, we use the project names as the keywords to search for potential vulnerabilities and collect vulnerability data from multiple sources, including vulnerability databases~\cite{nvd,mitre-cve}, security advisories~\cite{chrome-security,webkit-security,firefox-security}, bug-tracking systems~\cite{chromium-bug-tracker,webkit-bug-tracker,spidermonkey-bug-tracker}, GitHub repositories~\cite{v8-github,javascriptcore-github,spidermonkey-github,chakracore-github}, research blogs~\cite{project-zero-blog,v8-blog,mozilla-security-blog}, and conferences~\cite{blackhat,OffensiveCon2024,pwn2own,poc,zerocon}. 
% The detailed data resources are listed on our website~\cite{website}.

In total, we collected \textbf{1,215} candidate engine-related CVEs before further screening.
We first queried public CVE databases for each engine name.  
The terms \textit{v8} and \textit{chakracore} returned 325 and 177 CVEs, respectively.  
For JSC and SM the situation is different: their CVEs are usually reported under the browser projects that embed them (\textit{Safari} and \textit{Firefox}).\footnote{The keyword \emph{Safari} yields 1\,440 CVEs and \emph{Firefox} 3\,329 CVEs, the majority of which originate from non-JS components, including rendering engines, networking stacks, and sandbox.}  
To identify the engine-specific entries, we followed the CVE pages to their linked bug-tracking records, which list the affected component. This triage produced 635 SM CVEs. As many WebKit Bugzilla reports are private, the public CVE record for JSC is incomplete, as we are only able to collect 10 CVEs through standard sources. To compensate, we mined Google Project Zero’s public issue tracker~\cite{projectzeroissues}, ultimately obtaining 78 JSC CVEs. Then we conduct data screening to filter out vulnerabilities that are not out of date or lack sufficient information based on the following criteria:

\begin{itemize}[leftmargin=*]
    \item From the initial \textbf{1,215} engine-related CVEs, we first discarded \textbf{638} outdated entries (e.g., those published before 2017), and then removed additional cases lacking sufficient technical detail (e.g., patch links or analysis reports), leaving \textbf{371} well-documented candidates.
    \item Given the significant effort required for manual vulnerability analysis~\footnote{It is worth noting that analyzing a single vulnerability, including bug characteristic analysis, exploitation, and patch analysis, can be highly time-consuming, taking anywhere from several hours to days depending on its complexity.} and the constraints of a limited manual analysis budget, we prioritize severe vulnerabilities to balance the depth and breadth of our work. Specifically, we retain only those vulnerabilities with a CVSS score higher than 7.0 (classified as high risk) or those that have received bounty awards, which left us with \textbf{241} vulnerabilities.
\end{itemize}

After data screening, we obtain \textbf{241} JS engine vulnerabilities for subsequent manual analysis. We also collect the detailed patch analysis from corresponding bug-tracking systems to better understand the root cause and exploitation of these vulnerabilities. These systems include \textit{Chromium's bug tracker}~\cite{chromium-bug-tracker} for \textit{V8}, \textit{WebKit’s Bugzilla}~\cite{webkit-bug-tracker} for \textit{JSC}, \textit{Mozilla’s Bugzilla}~\cite{spidermonkey-bug-tracker} for \textit{SM}, and \textit{GitHub Issues}~\cite{chakracore-github} for \textit{CH}. From these systems, we obtain \textit{bug resolution histories} and additional contextual data regarding the vulnerabilities, such as \textit{associated exploit scripts} and \textit{developer comments}, which are invaluable for enriching our dataset with practical insights.

To further investigate the exploitation mechanisms for these vulnerabilities, we manually assess their exploitability and obtain \textbf{75} cases with exploitable proofs of concept for exploit triggering chain analysis. During our analysis, we additionally consult security blogs (e.g., Google Project Zero) and papers presented at security-related conferences (e.g., Black Hat and Offensive Con) to get \textit{detailed analysis reports} that provide in-depth exploit analyses, offering insights that are usually not available in official bug databases or advisories.

\subsection{Case Study: CVE-2019-13728}
\label{appendix:CVE-2019-13728}

\begin{figure}[t]
    \centering
    \includegraphics[width=0.9\linewidth]{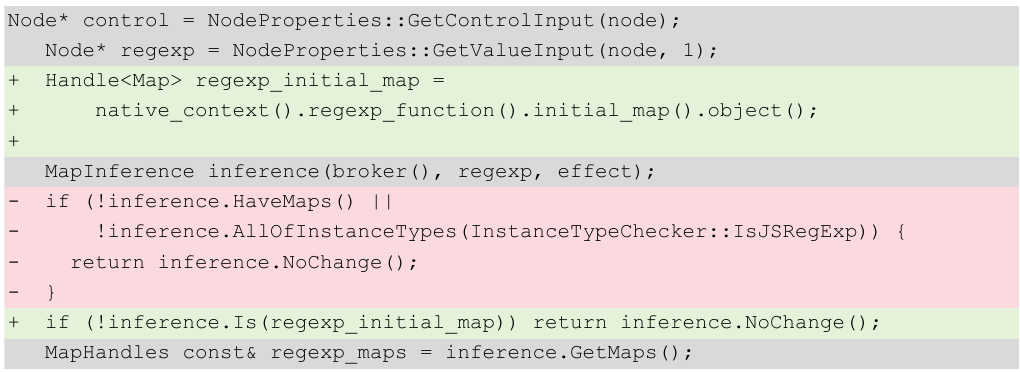}
    \captionsetup{justification=centering}
    % \caption{CVE-2019-13728: \textit{Incorrect Type Handling} vulnerability in V8}
    \caption{A vulnerability in V8 caused by \textit{Incorrect Type Handling} \#CVE-2019-13728}
    \label{fig:CVE-2019-13728}
\end{figure}

\autoref{fig:CVE-2019-13728} illustrates CVE-2019-13728~\cite{CVE-2019-13728}, 
which is caused by an incorrect type assumption in V8's handling of \textit{RegExp} objects. 
V8 incorrectly assumes that unmodified \textit{exec} methods imply the object retains its original map, 
checked only by \textit{InstanceType}. 
This causes JIT-compiled code to access properties at assumed offsets, 
resulting in Out-of-Bounds writes when the object's layout changes.

\subsection{Case Study: CVE-2017-0234}
\label{appendix:CVE-2017-0234}

\begin{figure}[t]
\centering
\includegraphics[width=0.9\linewidth]{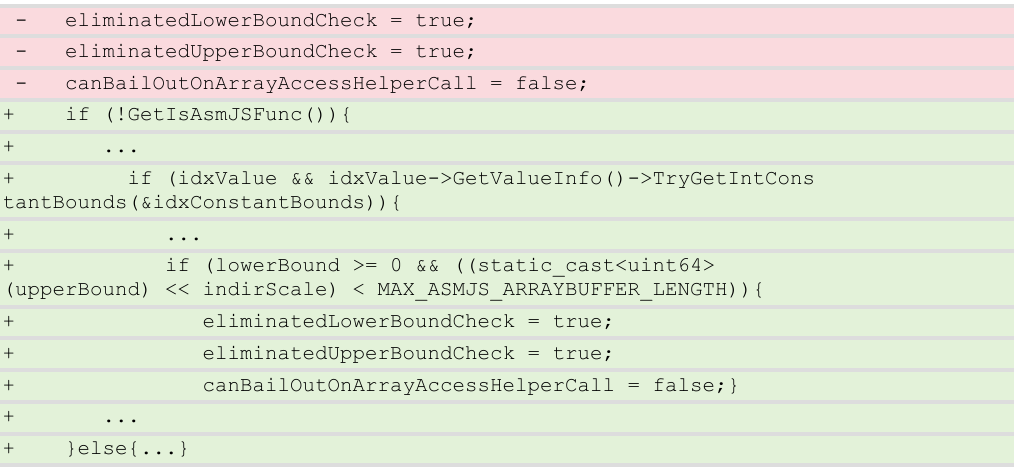}
\captionsetup{justification=centering}
\caption{A vulnerability in ChakraCore caused by \textit{Incorrect Optimization Algorithm} \#CVE-2017-0234}
\label{fig:CVE-2017-0234}
\end{figure}

\autoref{fig:CVE-2017-0234} illustrates CVE-2017-0234~\cite{CVE-2017-0234}, 
a representative vulnerability caused by an incorrect optimization algorithm in ChakraCore. 
This issue arises when the JIT compiler incorrectly removes boundary checks for virtual typed arrays, 
assuming that all array accesses remain within a 4 GB limit. 
Once this limit is exceeded, Out-of-Bounds writes can occur, potentially leading to memory corruption. 
The patch addresses this flaw by adding range checks to prevent unsafe boundary check eliminations.

\end{document}